\documentclass[aps,prd,twocolumn,nofootinbib]{revtex4-2}

\usepackage{booktabs,makecell,multirow}

\usepackage{graphicx}
\usepackage{float}
\usepackage{color}

\usepackage[colorlinks=true, linkcolor=blue, citecolor=blue, urlcolor=blue]{hyperref}

\newcommand*{\RMN}[1]{\uppercase\expandafter{\romannumeral#1}}

\begin{document}

\title{Updated Determination of Ellis-Jaffe Sum Rules up to $\rm N^{3}LO$ QCD corrections}	

\author{Hua Zhou$^{1}$}
\email{zhouhua@swust.edu.cn}

\author{Qing Yu$^{1}$}
\email{yuq@swust.edu.cn}

\author{Xing-Gang Wu$^{2}$}
\email{wuxg@cqu.edu.cn}

\affiliation{$^1$ School of Mathematics and Physics, Southwest University of Science and Technology, Mianyang 621010, China\\
$^2$ Department of Physics, Chongqing Key Laboratory for Strongly Coupled Physics, Chongqing University, Chongqing 401331, China}

\date{\today}

\begin{abstract}

In this paper, we explore the properties of the Ellis-Jaffe Sum Rule (EJSR) by employing the Principle of Maximum Conformality (PMC) approach to address its perturbative part up to next-to-next-to-next-to-leading order ($\rm N^{3}LO$) QCD contributions. By applying the PMC, we achieve a precise perturbative QCD prediction for the EJSR, free from conventional ambiguities associated with the renormalization scale choices. Considering the presence of the $\alpha_s$ Landau pole near the asymptotic scale, we incorporate the low-energy $\alpha_s$ model based on analytic perturbation theory (APT) to refine the EJSR behavior in the infrared region. By combining the PMC approach with the low-energy APT model, the agreement between theoretical calculations and experimental measurements of EJSR is significantly improved, as evidenced by the reduced discrepancy from $\chi^{2}/d.o. f|_{\rm Conv.}=1.86$ to $\chi^{2}/d.o. f|_{\rm PMC}=1.19$, thereby validating the effectiveness of our approach.

\end{abstract}

\maketitle

\section{Introduction}

The high-luminosity, precision collider will yield a wealth of high-precision experimental measurements about nucleon structure functions. This will further reveal the dynamic laws of strong interaction~\cite{Altarelli:1988nr, Efremov:1989sn, Ellis:1993te, Close:1993mv, Altarelli:1993np}. Moreover, these experimental advancements have also inspired numerous theoretical efforts to probe the structure functions and their first moment, i.e. the Ellis-Jaffe Sum Rule (EJSR)~\cite{Ellis:1973kp}. As a pivotal sum rule in particle physics, the EJSR provides a framework for elucidating the intricate spin structure in protons and neutrons, encompassing the complicated interplay of contributions from quarks and gluons, offering invaluable insights into their fundamental properties. A detailed examination of EJSR will enhance our comprehension of the fundamentals of strong interaction and Quantum Chromodynamics (QCD). The EJSR has been utilized to explore the hypothesis that strange quarks contribute minimally to the spin structure of nucleon~\cite{Deur:2018roz}. Precise determination of EJSR can also aid us in verifying the exact value of the Bjorken Sum Rule~\cite{Bjorken:1966jh, Kodaira:1979ib, Gorishnii:1985xm, Larin:1991tj, Braun:1986ty, Balitsky:1989jb, Yu:2021ofs}, thereby extracting the accurate value of the strong coupling constant $\alpha_s$.

The EJSR can be expressed as
\begin{eqnarray}
	M_1(Q)&=&\int^{1}_{0}dx g^{p(n)}_{1}(x,Q)\nonumber\\
	~&=&\frac{1}{36}C^{\rm ns}(Q,\mu_r)(\pm 3 g_A + a_8) \nonumber\\
	~& & + \frac{1}{9} C^{\rm s}(Q,\mu_r) a_0(Q),
\label{eq1}
\end{eqnarray}
where $g^{p(n)}_{1}(x,Q)$ represents the spin-dependent structure function of proton (neutron), incorporating the Bjorken scaling variable $x$. $\mu_r$ stands for the renormalization scale. $g_A$ represents the axial vector coupling constant, where the plus or minus sign in front of $g_A$ corresponds to the case of proton or neutron, respectively. $a_8$ is the isovector and flavor-octet axial charges of the nucleon~\cite{Close:1993mv} and $a_0(Q)$ is the flavor-singlet axial charge. Since the non-singlet axial current is conserved in the massless quark limit, this implies that the elements $g_A$ and $a_8$ are renormalization group invariants. $C^{\rm ns}$ and $C^{\rm s}$ are non-singlet and singlet coefficient functions, respectively, both of which are perturbatively calculable and exhibit a dependency on the renormalization scale for a fixed-order series.

Till now, $C^{\rm ns}$ and $C^{\rm s}$ have been calculated up to three-loop QCD calculations~\cite{Kodaira:1979pa, Larin:1994dr, Larin:1997qq}. However, the results still struggle to meet the expected accuracy standards, and there is a significant deviation from the experimental measurements. Substantial exploration and attempts have been undertaken to enhance the precision of theoretical prediction of the EJSR, such as the MSR scheme~\cite{Hoang:2009yr}, which aims to optimize the coefficient convergence of the perturbative expansion of the EJSR. In addition, the EJSR results also reveal significant renormalization scale uncertainty, which undoubtedly presents a challenge on our path toward achieving higher precision.

In conventional perturbation calculations, it is customary to set the renormalization scale $\mu_r=Q$ to eliminate large logarithmic terms $\log(\mu^2_r/Q^2)$. Subsequently, the renormalization scale is varied within a specified range to evaluate its uncertainty. However, this straightforward approach fails to satisfy the requirement of renormalization group invariance (RGI) and causes ambiguities in the renormalization scale and scheme~\cite{StueckelbergdeBreidenbach:1952pwl, Gell-Mann:1954yli, Callan:1970yg, Symanzik:1970rt, Peterman:1978tb}. According to the RGI, physical quantities should remain invariant irrespective of different renormalization scheme and scale choices. If computations are carried out to a sufficiently high order, the RGI will automatically be satisfied due to mutual cancellation of the perturbative expansion dependence on the renormalization scheme and scale at each orders. However, when dealing with limited fixed-order series, the RGI cannot be naturally achieved. More explicitly, if simply setting an arbitrary choice of renormalization scale to the initial perturbative QCD (pQCD) series, the mismatches between the strong coupling constant $\alpha_s$ at each order and its corresponding perturbative coefficient will lead to uncertainties in both the renormalization scale and scheme~\cite{Wu:2013ei, Wu:2014iba, Wu:2019mky}.

To put it differently, perturbative expansions at fixed order inherently contain uncertainties associated with the scheme and scale. An ill-suited scale selection will greatly reduce the accuracy of theoretical prediction, ultimately leading to substantial discrepancies between theoretical predictions and experimental results. Owing to the intricate nature of Feynman diagram calculations, only relatively low-order computations can currently be achieved. As a result, procuring a theoretical prediction for the EJSR that remains independent of the renormalization scale at finite orders is important for shedding light on nucleon structure functions. In this paper, we utilize the Principle of Maximum Conformality (PMC)~\cite{Brodsky:2012rj, Brodsky:2011ta, Brodsky:2011ig, Mojaza:2012mf, Brodsky:2013vpa} approach to obtain a theoretical prediction of the EJSR that is independent of the renormalization scale.

The running behavior of $\alpha_s$ is controlled by the renormalization group equation (RGE),
\begin{equation}
	\frac{d\alpha_{s}(\mu_r)}{d\ln\mu^2_r}=\beta(\alpha_s)=-\sum^{\infty}_{i=0}\beta_{i}\alpha_{s}^{i+2}(\mu_r).
	\label{eq2}
\end{equation}
The $\{\beta_{i}\}$-functions have been computed up to 5-loops under the modified minimal-subtraction ($\overline{\rm MS}$) scheme~\cite{Baikov:2016tgj, Luthe:2016ima, Herzog:2017ohr, Chetyrkin:2017bjc}. The $\{\beta_{i}\}$-terms that appear in the pQCD series can be reabsorbed into $\alpha_s$, thus facilitating an accurate determination of the correct value of $\alpha_s$. Based on this, the PMC is considered as a strict and systematic approach to eliminate the ambiguities inherent in the renormalization scheme and scale. To be specific, the PMC single-scale setting approach determines a global effective coupling $\alpha_s(Q_\ast)$~\cite{Shen:2017pdu, Yan:2022foz} (where $Q_\ast$ represents the PMC scale, which serves as a proxy for the effective momentum flow within the process.) by utilizing all the RGE-involved non-conformal $\{\beta_i\}$ terms. Simultaneously, by absorbing all the non-conformal terms in the perturbative expressions, the PMC prediction is also independent of the choice of renormalization scheme~\cite{Wu:2018cmb}, and meets the basic requirement of RGI of physical observables.

In this paper, we aim to employ the PMC single-scale setting approach to attain a renormalization scale-invariant prediction for EJSR. Under the $\overline{\rm MS}$ scheme, the renormalized EJSR up to next-to-next-to-next-to-leading order (N$^{3}$LO) QCD corrections~\cite{Larin:1997qq} can be expressed in the expansion of strong coupling constant,
\begin{eqnarray}
M_1(Q)&=& \sum_{i=0}^{3} r_{i} a^{i}_{s}(\mu_r) + \cdots,
\end{eqnarray}
where $a_s(\mu_r)=\alpha_{s}(\mu_r)/\pi$. The coefficients $r_{i}$ and the followed $r_{i,j}$ are generally scale dependent, and for convenience we do not write out their scale dependence in the formulas throughout the paper. It is found that the renormalization scale dependence starts at the NNLO level, leading to considerable conventional scale uncertainties. To address this, the PMC approach will be adopted to eliminate the uncertainties stemming from the renormalization scale, thereby enabling more precise predictions for EJSR.

\section{CALCULATION TECHNOLOGY}

Using the degenerate relationship~\cite{Bi:2015wea}, the EJSR can be rewritten in the following perturbative form:
\begin{eqnarray}
M_1(Q)&=&r_{0,0}+r_{1,0}a_s(\mu_r)+(r_{2,0}+\beta_{0}r_{2,1})a^{2}_s(\mu_r)+ \nonumber\\
& & (r_{3,0}+2\beta_{0}r_{3,1}+\beta^2_0r_{3,2}+\beta_{1}r_{2,1})a^3_s(\mu_r)+\cdots,
\end{eqnarray}
where
\begin{eqnarray}
r_{0}&=&r_{0,0},\\
r_{1}&=&r_{1,0},\\
r_{2}&=&r_{2,0}+\beta_{0}r_{2,1},\\
r_{3}&=&r_{3,0}+\beta_{1}r_{2,1}+2\beta_{0}r_{3,1}+\beta^{2}_{0}r_{3,2}.
\end{eqnarray}
Inversely, the coefficients $r_{i,j}$ can be derived from the $n_f$-series in the coefficients $r_{i}$ given in Refs.~\cite{Larin:1997qq, Kodaira:1979pa, Larin:1994dr}, where $n_f$ is the number of light flavors, e.g.,
\begin{eqnarray}
r_{0,0}&=&0.138,\\
r_{1,0}&=&-0.127,\\
r_{2,0}&=&0.103,\\
r_{2,1}&=&-0.244-0.400\log\left(\frac{\mu_r}{Q}\right), \\
r_{3,0}&=&2.665,\\
r_{3,1}&=&-0.061-1.710\log\left(\frac{\mu_r}{Q}\right), \\
r_{3,2}&=&-0.780-1.535\log\left(\frac{\mu_r}{Q}\right).
\end{eqnarray}
The scale-invariant conformal coefficients at each orders are represented by $r_{i,0}$, and the non-conformal coefficients are denoted as $r_{i,j}$ ($j\neq 0$):
\begin{eqnarray}
	r_{i,j}&=&\sum^{j}_{k=0} C^{k}_{j} \ln^{k}(\mu^2_r /Q^2)\hat{r}_{i-k,j-k},
\end{eqnarray}
where $\hat{r}_{i,j}=r_{i,j}|_{\mu_r=Q}$ and $C^{k}_{j}=j!/k!(j-k)!$. After applying standard PMC procedure~\cite{Shen:2017pdu, Yan:2022foz}, all non-conformal $\{\beta_i\}$-terms are absorbed into $\alpha_s$, resulting in a conformal series that is both renormalization scale and renormalization scheme independent, e.g.,
\begin{eqnarray}
M_1(Q)&=& \sum_{i=0}^{3} \hat{r}_{i,0}a^{i}_s(Q_{\ast})+\cdots.
\end{eqnarray}
Leveraging the known perturbation series up to N$^3$LO-level, the PMC scale can be determined at the next-to-leading log (NLL) accuracy:
\begin{eqnarray}
\ln\frac{Q^2_\ast}{Q^2}&=&T_0 +T_1 a_s(Q^2),
\label{eqq}
\end{eqnarray}
where
\begin{eqnarray}
    T_{0} &=& -\frac{\hat r_{2,1}}{ r_{1,0}},  \\
    T_{1} &=& \frac{2(\hat{r}_{2,0}\hat{r}_{2,1}-\hat{r}_{1,0}\hat{r}_{3,1})}{\hat{r}_{1,0}}+\frac{\hat{r}^{2}_{2,1}-\hat{r}_{1,0}\hat{r}_{3,2}}{\hat{r}_{1,0}}\beta_{0}.
\end{eqnarray}
Notably, $Q_\ast$ remains uninfluenced by any selection of renormalization scale $\mu_r$. Thus, the ambiguity associated with the renormalization scale is effectively eliminated, thus improving the precision of EJSR.

\section{NUMERICAL RESULTS AND DISCUSSIONS}

For numerical calculation, we use $g_A=1.2723\pm0.0023$ \cite{ParticleDataGroup:2018ovx} and $a_8 = 0.58\pm0.03$ \cite{Close:1993mv}. To facilitate ease of understanding and manipulation, we defined $a_0(Q)$ in a proper invariant way as a constant, e.g., $a_0=0.141$ at $Q=5$ GeV for further discussion \cite{Hoang:2009yr}.

\subsection{The strong coupling constant $\alpha_s$}

\begin{figure}[htb]
\centering
\includegraphics[width=0.48\textwidth]{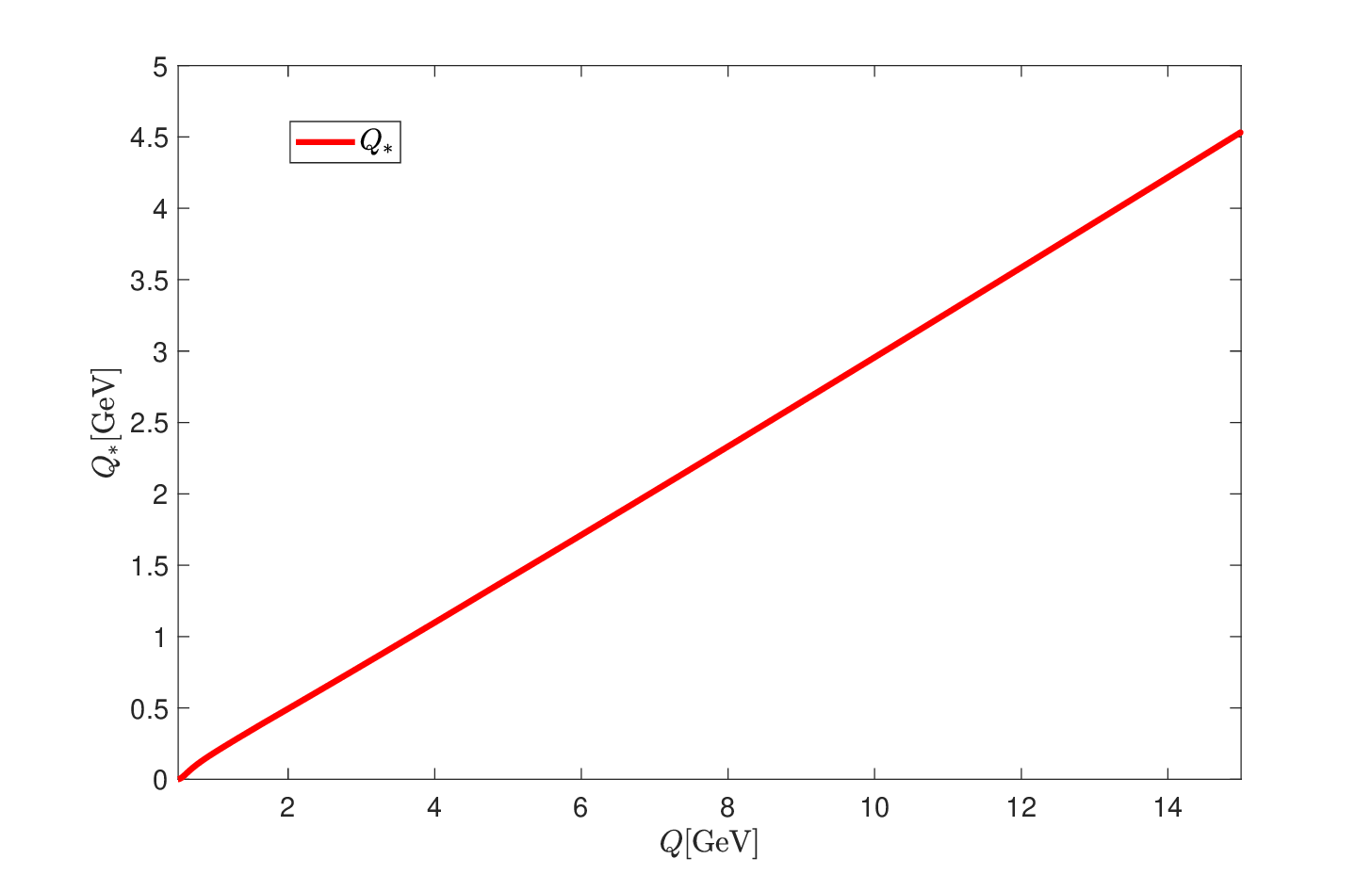}
\caption{The variation of $Q_\ast$ versus energy scale $Q$.}
\label{figQS}
\end{figure}

Using Eq.(\ref{eqq}), we can derive the relationship between $Q_\ast$ and $Q$, which is shown in Fig.\ref{figQS}. It shows that when $Q$ is small, the effective momentum flow $Q_\ast$ approaches the asymptotic scale $\Lambda$. The strong coupling constant $\alpha_s$ has an unreasonable maximum near the asymptotic scale, i.e., the Landau pole problem in this energy range will make it difficult to obtain a convincing pQCD prediction. To redefine the infrared behavior of $\alpha_s$ within small energy region, and to achieve reliable theoretical predictions, non-perturbative methods are typically employed to capture the intricate physics occurring in low energy region. Those methods include some phenomenological low-energy models, the lattice QCD, and the Dyson-Schwinger equations~\cite{Yu:2021yvw}.

For the EJSR to be addressed in this paper, low-energy models are adequate to describe the physics in these energy ranges and yield highly accurate theoretical predictions. Therefore, in our investigation of the running behavior of $\alpha_s$ at low energies, we primarily utilize low-energy model methods to enhance the precision of perturbative theoretical predictions. In the literature, numerous low-energy models for $\alpha_s$ have been proposed \cite{Cornwall:1981zr, Godfrey:1985xj, Halzen:1992vd, Shirkov:1997wi, Webber:1998um, Badalian:2001by, Brodsky:2010ur, Shirkov:2004ar, Shirkov:1999hm, Shirkov:2012ux}. In this paper, we adopt the analytical perturbation theory (APT) model \cite{Shirkov:1997wi} as a typical framework for redefining the infrared behavior of $\alpha_s$.

The APT model, analogous to the analysis of the general expression of effective charge $\alpha_s$ in QED, employs the perturbative analytical expression of the spectral function to address the Landau pole issue. Based on the expression of the one-loop $\alpha_s$, its corresponding spectral function can be briefly expressed as:
\begin{eqnarray}
	\rho(\delta)=\frac{\pi \beta^{-1}_0}{(\ln\frac{\delta}{\Lambda^2})^2+\pi^2}.
	\label{eq10}
\end{eqnarray}
The analytic running coupling is
\begin{eqnarray}
	\alpha_{an}(Q)=\frac{1}{\pi}\int^{\infty}_{0} \frac{d\delta \rho(\delta)}{\delta+Q^2}.
	\label{eq11}
\end{eqnarray}
After reconstruction, the space-like domain of $\alpha_s$ effectively eliminates all non-physical singularities, and the integration of spectral functions does not require additional subtraction. Based on Eqs.(\ref{eq10}, \ref{eq11}), the effective coupling of the APT model is formulated as
\begin{eqnarray}
	\alpha^{\rm APT}_{s}(Q)=\frac{\pi}{\beta_0}(\frac{1}{\ln k}+\frac{1}{1-k}),
\end{eqnarray}
where $k=Q^2/\Lambda^2$. Furthermore, it is worth mentioning that $\alpha^{\rm APT}_s(Q)$ does not incorporate any modified parameters; instead, it reconstructs the expression. The scale $\Lambda$ can be reformulated in the following manner:
\begin{eqnarray}
	\Lambda^2=Q^2 e^{-\phi(\frac{\beta_0 \alpha_s(Q)}{\pi})},
\end{eqnarray}
and $\phi(z)$ satisfies the evolution equation
\begin{eqnarray}
 \frac{1}{1-e^{\phi(z)}}+\frac{1}{\phi(z)}=z.
\end{eqnarray}

The APT model has been extended up to the three-loop accuracy~\cite{Magradze:2000hz, Kourashev:2001kd, Magradze:2005ab, Bakulev:2005gw, Bakulev:2006ex, Gabdrakhmanov:2025afi}; in this paper, we employ the \texttt{anQCD} Mathematica package~\cite{Ayala:2014pha} for numerical determination of the strong coupling constant $\alpha_s$ under the three-loop accuracy APT model.

\begin{figure}[htb]
\centering
\includegraphics[width=0.48\textwidth]{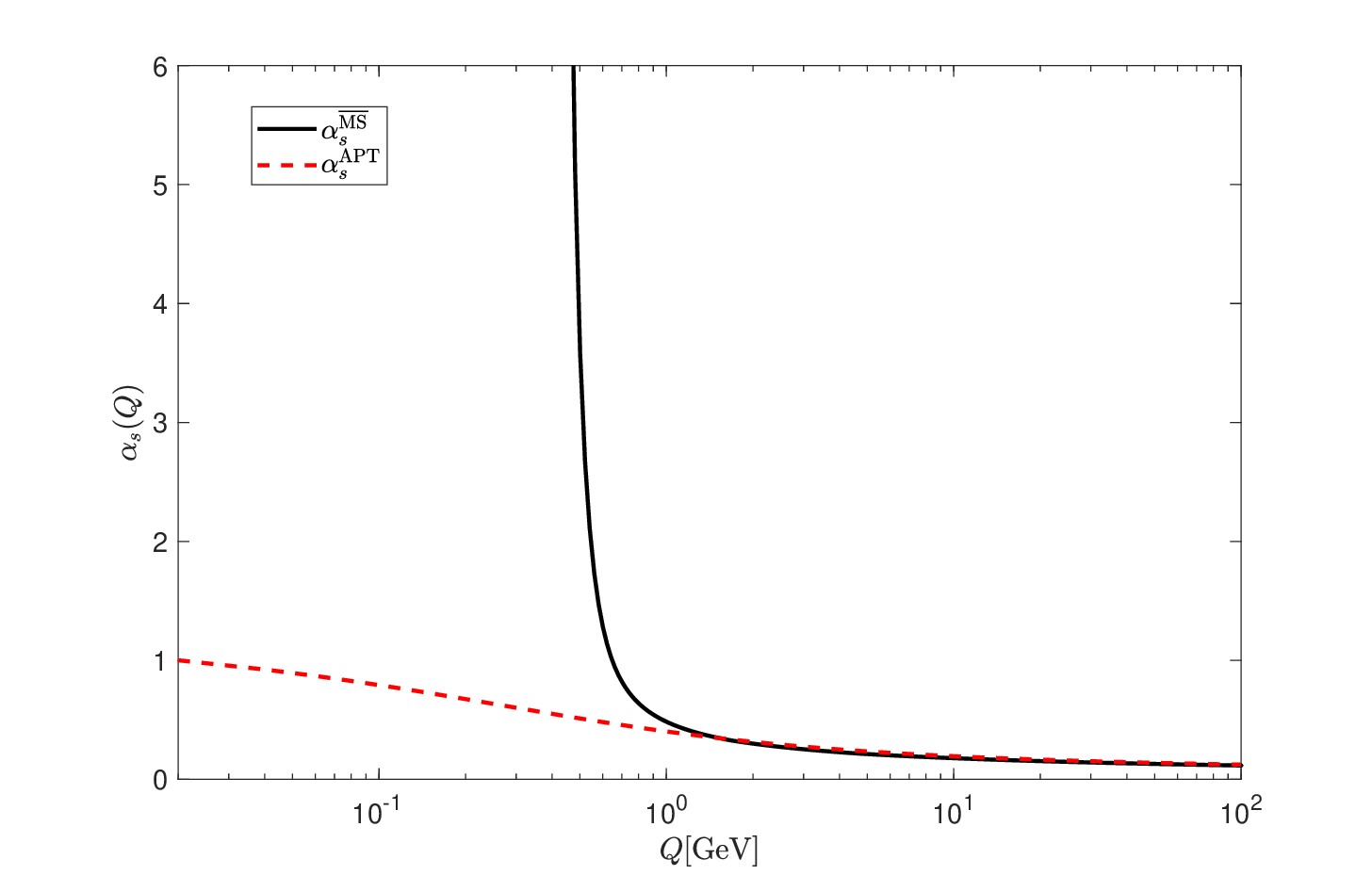}
\caption{The strong coupling $\alpha_s(Q)$ varies with the energy scale $Q$. The dotted line represents the three-loop APT low-energy model, while the three-loop $\alpha_s$ running behavior under the $\overline{\rm MS}$-scheme is depicted as a solid line.}
\label{figle}
\end{figure}

Fig.\ref{figle} depicts the scale running of $\alpha_s$, where the dotted line represents the three-loop APT running and the solid line denotes the solution of three-loop RGE (\ref{eq2}). A smooth transition of $\alpha_s$ between the low and high-energy regions is achieved by implementing the matching scheme proposed in Ref.\cite{Deur:2014qfa}, i.e., the transition scale for $\alpha_s$ is definitively set at $Q_0 \approx 1.535~\rm GeV$ by enforcing the condition that the first derivatives of $\alpha_s$ coincide at the intersection point of those two distinct energy regions.

\subsection{The $\rm N^3LO$-level EJSR $M_1(Q)$ together with an estimation of $\rm N^4LO$ contributions}

\begin{table*}[htb]
\renewcommand{\arraystretch}{1.5}
\begin{center}
\begin{tabular}{c c c  c  c c }
\hline
 ~~~$Q=6$~GeV~~~ & ~~~LO~~~ & ~~~NLO~~~ & ~~~$\rm N^{2}LO$~~~ & ~~~$\rm N^{3}LO$~~~ &  ~~~Total~~~ \\
\hline
 $\rm Conv.$ & $0.1378$ & $-0.0090^{-0.0021}_{+0.0015}$  & $-0.0022^{+0.0036}_{-0.0015}$ & $-0.0010^{+0.0045}_{-0.0013}$ &  $0.1256^{+0.0060}_{-0.0013}$\\
 $\rm PMC$ & $0.1378$ & $-0.0137$ & $-0.0012$ & $0.0033$ & $0.1286$\\
\hline
\end{tabular}
\caption{Total $\rm N^3LO$ value and separate values of each order terms for the EJSR $M_{1}(Q=6~{\rm GeV})$ under conventional and PMC scale-setting methods, respectively. For conventional prediction, the central value is for $\mu_r = Q$, while the upper and lower error correspond to variations over $\mu_r \in [Q/2, 2Q]$.}
\label{tabconv}
\end{center}
\end{table*}

Table \ref{tabconv} compares the total and the order-by-order contributions to $M_{1}(Q)$ calculated using conventional (Conv.) and PMC scale-setting approaches, respectively. The case of intermediate energy scale with $Q=6$ GeV has been adopted as an explanation. When setting the renormalization scale $\mu_r=Q$ for conventional method, the magnitude of $\rm N^{3}LO$-terms shows better convergence than that of the PMC method. However, this result emerges as an artefact of the logarithmic terms elimination, each terms of the initial pQCD series are highly scale-dependent. Specifically, when accounting for the renormalization scale uncertainties, the conventional series exhibits significant variations, e.g., the magnitude of $\rm N^{3}LO$-terms is $-0.0010^{+0.0045}_{-0.0013}$ for $\mu_r\in[Q/2,2Q]$. Consequently, the coefficient convergence in the conventional approach is less stable than that predicted by the PMC method. Thus, to have a scale-invariant series, which can be treated as the intrinsic perturbative behavior of the series, is important. Moreover, it is also important to have an estimation of the uncalculated higher-order (UHO) $\rm N^{4}LO$ contribution. It will be found that the scale-invariant PMC series can also obtain a more reliable and accurate estimation of UHO terms. 

At present, many attempts have been made in the literature, such as to propose a proper generating function or to construct a proper probability distribution and etc., however there is still no reliable method that is widely accepted to predict the UHO-contributions. At present, we employ the Pad$\acute{e}$ approximation approach (PAA)~\cite{Basdevant:1972fe, Samuel:1992qg, Samuel:1995jc} to have an estimation of the uncalculated $\rm N^{4}LO$-terms and to further assess the convergence of the perturbative behavior of EJSR series. According to our previous experiences, the PAA works effectively when enough higher-order terms are known. Within the PAA, a general up to $n_{\rm th}$-order pQCD series $\rho_{n}$ can be written as a fractional form, e.g. the following $[N/M]$-type generating function,
\begin{eqnarray}
\rho^{[N/M]}_{n} &=&a^{p}_s\times\frac{b_{0}+b_{1}a_s + \cdots +b_{N}a^{N}_s}{1+c_{1}+\cdots c_{M}a^{M}_s},\\
&=&\sum^{n}_{i=1} C_{i}a^{p+i-1}_{s} +C_{n+1}a^{p+n}_{s} +\cdots,
\end{eqnarray}
where $p=0$ for the present case of EJSR, which indicates the leading-order terms is a constant free of $\alpha_s$. The expansion parameters $b_{i\in[0,N]}$ and $c_{i\in[1,M]}$ can be systematically determined from the known perturbative coefficients $C_{i\in[1,n]}$. The first unknown coefficient $C_{n+1}$ becomes expressible through these determined parameters $b_{i\in[0,N]}$ and $c_{i\in[1,M]}$ thereby establishing its dependence on the complete set of known coefficients $\{C_{1},...,C_{n}\}$.

Using the current $\rm N^{3}LO$-level EJSR series, we will adopt the $[0/2]$-type PAA for estimating the uncalculated $\rm N^{4}LO$-terms~\cite{Shen:2016dnq, Du:2018dma}~\footnote{This type of PAA agrees with the GM-L procedure~\cite{Gell-Mann:1954yli} to obtain scale-independent perturbative QED predictions and is consistent with the Generalized Crewther Relations~\cite{Shen:2016dnq}.}, which gives
\begin{eqnarray}
\Delta M_{1}(Q)|_{\rm N^{4}LO,\; \rm Conv.} &=&\frac{r^{3}_{2} + 2 r_1 r_2 r_3}{r^{2}_{1}},\\
\Delta M_{1}(Q)|_{\rm N^{4}LO,\; \rm PMC} &=&\frac{r^{3}_{2,0}+2r_{1,0} r_{2,0} r_{3,0}}{r^{2}_{1,0}}.
\end{eqnarray}
Numerically, the $\rm N^{4}LO$-terms' contribution predicted by the $[0/2]$-type PAA at $Q=6$ GeV under conventional and PMC methods are

\begin{eqnarray}
\Delta M_{1}(Q)|_{\rm N^{4}LO,\; \rm Conv.} &=&-0.0006^{-0.0003}_{-0.0025},\\
\Delta M_{1}(Q)|_{\rm N^{4}LO,\; \rm PMC} &=&-0.0006,
\end{eqnarray}
where PMC prediction is scale-invariant and the error of conventional series is from $\mu_r\in[Q/2,2Q]$.

\begin{figure}[htb]
\centering
\includegraphics[width=0.48\textwidth]{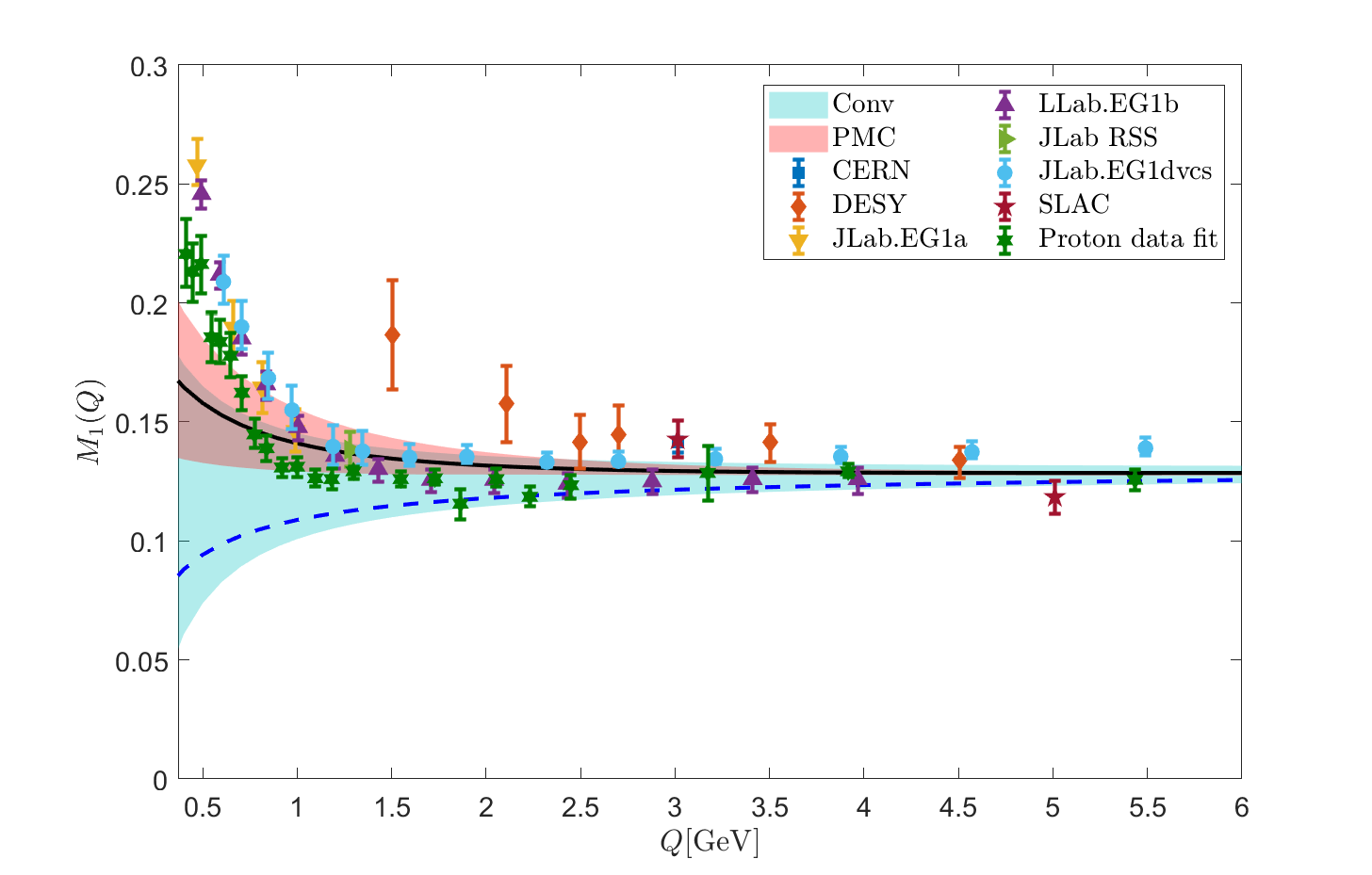}
\caption{The $Q$-dependent EJSR derived under conventional and PMC methods, respectively. The error of PMC series comes from the estimated magnitude of the uncalculated $\rm N^{4}LO$-terms, while the error of conventional series comes from the combined error of the estimated magnitude of the uncalculated $\rm N^{4}LO$-terms and $\mu_r\in[Q/2,2Q]$.}
\label{figf}
\end{figure}

The magnitudes of the uncalculated $\rm N^{4}LO$-terms is incorporated as a component of theoretical uncertainties in our error analysis, as presented in Fig.\ref{figf}, which shows the comparison of the EJSR obtained through both the PMC and the conventional methods with the experimental measures of various groups. The solid line corresponds to the PMC-determined EJSR center value, and its error band incorporates the uncertainties from the estimated magnitude of the $\rm N^{4}LO$-terms. The dashed line stands for conventional prediction of EJSR center value, and its error band denotes the combined uncertainties originating from renormalization scale variations $\mu_r\in[Q/2,2Q]$ and the estimated magnitude of the $\rm N^{4}LO$-terms. Conservatively, we have implicitly adopt $\pm \left|\Delta M_{1}(Q)\right|_{\rm N^{4}LO}$ in Fig.\ref{figf} as an estimation of the error caused by uncalculated $\rm N^{4}LO$-terms for both cases.

The colorful error bars represent the experimental and fitting results outcomes reported in Refs.\cite{Osipenko:2005nx, CLAS:2017qga}. The results demonstrate that the PMC approach eliminates renormalization scale uncertainties inherent in conventional calculations, thereby enabling more precise theoretical predictions. For the EJSR observable, PMC exhibits superior performance compared to conventional methods: it eliminates renormalization scale uncertainties while achieving smaller theoretical errors. Specifically, at $Q=2$ GeV, conventional predictions show a peak deviation of 15.06\% from the central value, whereas the PMC suppresses these uncertainties to 4.27\% -- the error has decreased to one-fourth of its conventional results -- while showing better agreement with experimental data.

We then employ the $\chi^{2}/d.o.f$ to quantitatively evaluate the degree of concordance between theoretical predictions and experimental data. The $\chi^{2}/d.o.f$ assesses the consistency between the theory-predicted EJSR values and their experimental counterparts,
\begin{equation}
\chi^{2}/d.o.f = \frac{1}{N}\sum_{j=1}^{N} \left[\frac{M_{1}(Q_j)|_{\rm exp.}-M_{1}(Q_j)|_{\rm the.}}{\sigma_{j}}\right]^{2},
\end{equation}
where ``exp." represents the experimental value and ``the." refers to the central value of the theoretical prediction. $\sigma_{j}$ represents the combined uncertainty encompassing both the experimental measurement error of the $j$-th data point and the corresponding theoretical prediction error. The Refs.\cite{Osipenko:2005nx, CLAS:2017qga} show the number of data points $N=67$. It is observed that by utilizing the PMC to enhance the precision of pQCD contribution, theoretical prediction shows a much better agreement with the data, e.g., the application of the PMC yields a significantly smaller $\chi^{2}/d.o.f|_{\rm PMC}=1.19$ compared to the conventional method $\chi^{2}/d.o.f|_{\rm Conv.}=1.86$.

Practically, the Pearson correlation coefficient (PCC) has also been adopted to roughly evaluate the consistency between theoretical predictions and experimental results, and the PCC expression can be written as
\begin{eqnarray}
	R_{\rm PCC}&=&\frac{\sum (y^{\rm exp.}_{i}-\overline{y}^{\rm exp.})(y^{\rm the.}_{i}-\overline{y}^{\rm the.})}{\sqrt{\sum (y^{\rm exp.}_{i}-\overline{y}^{\rm exp.})^{2}\sum(y^{\rm the.}_{i}-\overline{y}^{\rm the.})^{2}}},
\end{eqnarray}
where $R_{\rm PCC}\approx 1$ indicates a strong positive correlation with a good fit, $R_{\rm PCC}\approx 0$ indicates no correlation and a poor fit and $R_{\rm PCC}\approx -1$ indicates a negative correlation. For the present case of EJSR, we have $R_{\rm PCC}|_{\rm PMC}=0.862$ while $R_{\rm PCC}|_{\rm Conv.}=-0.820$. This also suggests that the PMC series can achieve a much better agreement with the data.

\begin{figure}[htb]
\centering
\includegraphics[width=0.48\textwidth]{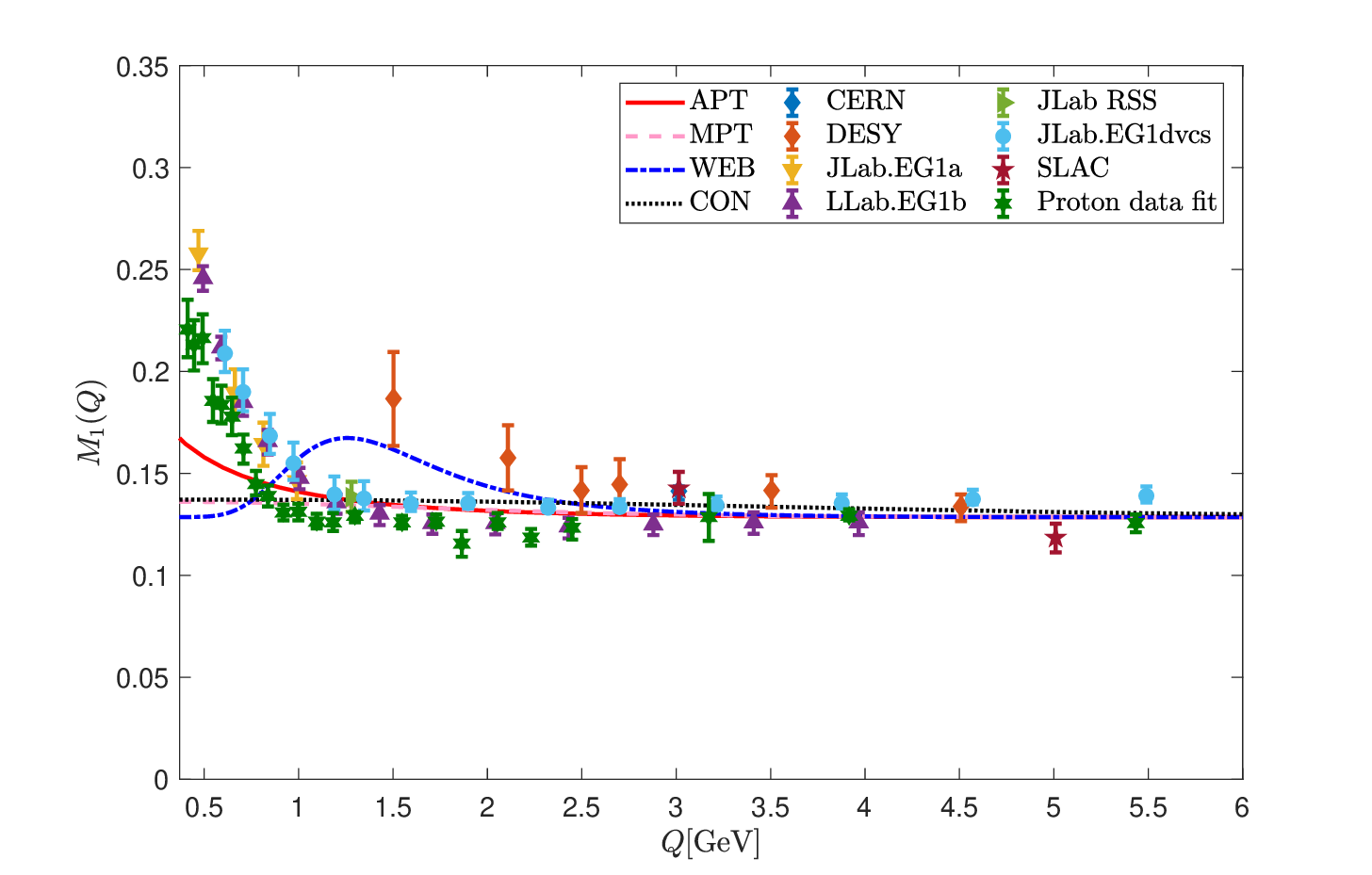}
\caption{A comparison of the PMC predictions on the EJSR by using four low-energy models, e.g., APT, CON, MPT, WEB~\cite{Halzen:1992vd, Shirkov:1997wi, Shirkov:1999hm, Shirkov:2012ux, Webber:1998um}.}
\label{fig4}
\end{figure}

As a final remark, in our above calculation, we have adopted the APT model to address the infrared divergence behavior of $\alpha_s$. In Fig.\ref{fig4}, we give a comparison of how the EJSR is affected by various low-energy models, such as APT, the continuum QCD theory (CON), the massive pQCD model (MPT), and the Webber model (WEB)~\cite{Halzen:1992vd, Shirkov:1997wi, Shirkov:1999hm, Shirkov:2012ux, Webber:1998um}. In this comparison, the CON and WEB low-energy models adopted the running coupling with one-loop accuracy, while the MPT low-energy model incorporates the three-loop accuracy for the subsequent discussion. Definitions and a comparison of various low-energy models can be found in Ref.\cite{Zhang:2014qqa}. The results derived by taking the input parameters to be their central values are presented in Fig.\ref{fig4}. It shows that the APT model does provide a better explanation of data.

\section{SUMMARY}

As a summary, in this paper, we have employed the PMC single-scale setting method to address the pQCD calculable contribution to the EJSR with an accuracy up to ${\rm N^{3}LO}$. The possible ${\rm N^{4}LO}$ contributions to the EJSR have also been estimated by using the PAA. Our analyses show that the conventional renormalization scale ambiguity can be removed by applying the PMC and the precision of the EJSR can be eliminated, thereby the net theoretical uncertainty is significantly reduced. Furthermore, by using the low-energy APT model, we can achieve a more stable pQCD approximant for the EJSR. As shown by Fig.~\ref{figf}, the EJSR result obtained using PMC aligns more closely with the experimental data, and the theoretical-experiment consistency is optimized from $\chi^{2}/d.o.f|_{\rm Conv.}=1.86$ to $\chi^{2}/d.o.f|_{\rm PMC}=1.19$, thereby validating the effectiveness of our approach. The application of the PCC approach further validates that the PMC series exhibits significantly enhanced predictive consistency with experimental data. Thus our present results emphasize the necessity of a proper renormalization scale-setting method during the pQCD calculation.

\section{Acknowledgments}

This work was supported by the Natural Science Foundation of China under Grant No.12305091, No.12175025 and No.12347101, the Natural Science Foundation of Sichuan Province under Grant No.2024NSFSC1367, and the Research Fund for the Doctoral Program of the Southwest University of Science and Technology under Contract No.24zx7117 and No.23zx7122.

\end{document}